\RequirePackage{fix-cm}
\documentclass[smallextended]{svjour3}       

\smartqed  
\usepackage[pdftex]{graphicx}

\usepackage{amssymb}
\usepackage{amsmath}

\usepackage{color}

\newcommand{\RB}[1]{\left( #1 \right)}
\newcommand{\BR}[1]{\left[ #1 \right]}

\newcommand{\AV}[1]{\left< #1 \right>}
\newcommand{\PT}[1]{\frac{\partial #1}{\partial t}}

\newcommand{\BM}[1]{\mbox{\boldmath$\bf #1 $}}
\begin{document}

\title{Numerical Study on Entrance Length in Thermal Counterflow of Superfluid $^4$He
}


\author{Hiromichi Kobayashi \and Satoshi Yui \and Makoto Tsubota}


\institute{H. Kobayashi \at
              Department of Physics, Hiyoshi Campus, Keio University, Yokohama 223-8521, Japan \\
              Tel.: +81-45-566-1323\\
              \email{hkobayas@keio.jp}           
           \and
           S. Yui \and M. Tsubota \at
              Department of Physics, Osaka City University, Osaka 558-8585, Japan
}

\date{Received: date / Accepted: date}

\maketitle

\begin{abstract}
Three-dimensional numerical simulations in a square duct were conducted to investigate entrance lengths of normal fluid and superfluid flows in a thermal counterflow of superfluid $^4$He. The two fluids were coarse-grained by using the Hall-Vinen-Bekharevich-Khalatnikov (HVBK) model and were coupled through mutual friction. We solved the HVBK equations by parameterizing the coefficient of the mutual friction to consider the vortex line density. \textcolor{black}{A uniform mutual friction parameter was assumed in the streamwise direction.} Our simulation showed that the entrance length of the normal fluid from a hot end becomes shorter than that of a single normal fluid due to the mutual friction with the parabolically developed superfluid flow near the hot end. As the mutual friction increases, the entrance length decreases. Same as that, the entrance length of the superfluid from a cold end is affected by the strength of the mutual friction due to the parabolically developed normal fluid flow near the cold end. \textcolor{black}{Aside from the entrance effect, the realized condition of a tail-flattened flow is discussed by parameterizing the superfluid turbulent eddy viscosity and the mutual friction.} 
\end{abstract}

\vspace{-2mm}

\section{Introduction}
\label{sec:1}

\vspace{-1mm}

An inviscid fluid flow of $^4$He realizes below 2.17 K, 
and the superfluid $^4$He consists of the inviscid superfluid 
and viscous normal fluid components. 
The superfluidity is caused by Bose-Einstein condensation, 
and thus the superfluid vortex is quantized. 

In a duct flow filled with the superfluid $^4$He 
connected to a large helium bath, 
when the duct end opposed to the helium bath is heated, 
the normal fluid flow goes to the helium bath and 
the superfluid flow moves to the heated end 
to satisfy the total mass conservation. 
This experimental setup is termed as a thermal counterflow. 
As increasing a heat flux at the heated end, 
a relative velocity of the two fluids increases 
and mutual friction emerges in the thermal counterflow \cite{MF-exp}. 
The mutual friction is due to the tangle of the quantized vortex 
in the superfluid flow. 
The tangle structure was numerically demonstrated 
by using the vortex filament method (VFM) 
in a computational box under the periodic boundary condition \cite{Schwarz,Adachi}. 

A solid boundary effect was considered 
in flows between parallel walls \cite{Baggaley-Ldist} 
and in square duct flows \cite{Yui-duct}. 
In the simulations, although velocity profiles of the normal fluid flows 
were prescribed as parabolic, tail-flattened and turbulent flows, 
vortex line density distributions become bimodal 
in the wall-normal direction. 
Such a non-uniform distribution of the vortex line density 
was observed near the hot end in the experiments 
of the thermal counterflows \cite{Mantia-Ldist}. 
Recently, the VFM for the superfluid and 
Navier-Stokes equations for the normal fluid 
were fully coupled by the mutual friction 
and it was found that the vortex line density strength 
and its non-uniform distribution quite affect the velocity profiles 
of the two fluids \cite{PRL-Yui}. 

Entrance lengths of the normal fluid and the superfluid 
at the opposite pipe-ends of the thermal counterflows 
were studied by the Hall-Vinen-Bekharevich-Khalatnikov 
(HVBK) model \cite{HVBK}, 
and it was found that 
the entrance length of the normal fluid increases linearly 
with the Reynolds number 
whereas that of the superfluid is enhanced 
by up to one order of magnitude 
when compared to the normal fluid \cite{PRF-Roche}. 
At the cooled pipe end connected to the helium bath, 
a large recirculation region of the superfluid flow appears 
and the superfluid in the central region reenter 
into the helium bath. 
This flow configuration induces a tail-flattened flow 
of the normal fluid 
recently discovered in the duct flow experiments \cite{Marakov}. 

In this study, we investigate the influences of the mutual friction 
and its non-uniform distribution on the entrance lengths 
of the two fluids by means of the HVBK model. 
\textcolor{black}{Aside from the entrance effect, 
we will discuss the condition realizing the tail-flattened flow 
by parameterizing the superfluid turbulent eddy viscosity 
and the mutual friction.}

\vspace{-2mm}

\section{Governing equations and numerical methods}
\label{sec:2}

\vspace{-1mm}


We adopted the two-fluid model 
in which the superfluid $^4$He is composed of 
an inviscid superfluid component 
and a viscous normal fluid component. 
The density ratio of each component depends on temperature. 

We used the HVBK model 
that is the coarse-grained two-fluid model. 
The momentum equations are described as follows:
\begin{align}
\rho_n \BR{\PT{\BM{v}_n}+\RB{\BM{v}_n\cdot \nabla}\BM{v}_n} &= -\nabla p_n + \mu_n\nabla^2\BM{v}_n+\BM{F}_{ns}, \label{eq:vn} \\
\rho_s \BR{\PT{\BM{v}_s}+\RB{\BM{v}_s\cdot \nabla}\BM{v}_s} &= -\nabla p_s -\BM{F}_{ns} , \label{eq:vs}
\end{align}
where the subscripts of $n$ and $s$ denote the normal fluid and superfluid components, 
$\rho$ is the density, $\BM{v}$ is the velocity, $p$ is the pressure, 
$\mu$ is the molecular viscosity, and $\BM{F}_{ns}$ denote the mutual friction. 
The mutual friction is formulated as 
\begin{equation}
\BM{F}_{ns} = -a_f \RB{\BM{v}_n-\BM{v}_s}, \quad a_f = g \rho_s \kappa \alpha L ,
\label{eq:Fns}
\end{equation}
where $a_f$ is defined as a coefficient of the mutual friction, 
$g$ is the anisotropic parameter, 
$\kappa$ is the quantum circulation of the superfluid vortex, 
$\alpha$ is the coefficient as a function of temperature 
\textcolor{black}{\cite{rho-mu-B}}, 
and $L$ is the vortex line density \textcolor{black}{\cite{MF-exp,PRL-Yui}}. 
In the present study, 
we use $a_f$ as \textcolor{black}{an external} parameter changing the mutual friction, 
although \textcolor{black}{in real flow} $a_f$ is uniquely determined by 
the relative velocity $\BM{v}_n-\BM{v}_s$ and temperature 
that also determine the vortex line density $L$. 
The mutual friction depends on 
the vortex line density as shown 
in the two-fluid coupled simulation \cite{PRL-Yui}. 
Here, the superfluid $^4$He is treated as an incompressible flow, 
and continuity equations are adopted 
as $\nabla \cdot \BM{v}_n = \nabla \cdot \BM{v}_s =0$. 


\begin{figure}[t]
\centering
\includegraphics[width=0.95\linewidth]{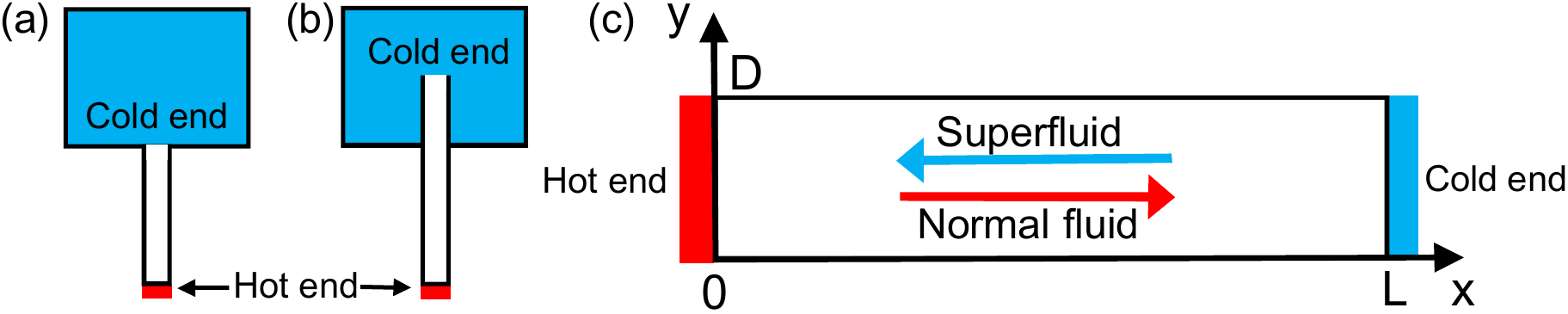}
\caption{Configurations (a) and (b) of the experimental setup of the duct for thermal counterflow, and (c) computational domain.}
\label{fig:domain}
\end{figure}

We adopted commonly used numerical methods 
for the incompressible flow. 
The second order finite difference method 
and the second order Adams-Bashforth method 
are utilized 
for the spatial discretization and temporal integration, respectively. 
The velocity and pressure are coupled by using the MAC 
\textcolor{black}{(marker and cell)} scheme \textcolor{black}{\cite{MAC}}, 
and the Poisson equation of pressure is solved 
by the BiCGSTAB \textcolor{black}{(biconjugate gradient stabilized)} method 
\textcolor{black}{\cite{BCG}}. 

Figure \ref{fig:domain} shows the configurations (a) and (b) 
of the experimental setup of the duct for thermal counterflow, 
and (c) computational domain. 
In the configuration (a), the duct end is connected 
just beneath the helium bath colored on blue. 
In this configuration, the normal fluid flow abruptly expands 
into the helium bath and it causes the strong recirculation 
of the superfluid flow. 
It is reported that 
the recirculation affects the entrance length near the cold end 
and the superfluid boundary layer \cite{PRF-Roche}; 
hereafter we call the reference \cite{PRF-Roche} 
as the BLR \textcolor{black}{(Bertolaccini, L\'ev\^eque and Roche)} study. 
By contrast, we assume the configuration (b) 
as shown in the recent experiment \cite{Marakov}. 
The duct end is inserted into the inside of the helium bath. 
\textcolor{black}{In the present study, 
we assume that the superfluid enters 
at the cold end without recirculation.} 

We conducted three-dimensional two-fluid simulations 
of the thermal counterflow in the duct filled with the superfluid $^4$He. 
The computational domain is the square duct of $L \times D \times D$, 
$L$ is the duct length and $D$ is the duct width and height 
in the cross section. We assume the hot end at $x = 0$ and 
the cold end at $x = L$. 
We chose the duct size of $D = 1$ cm and $L = 30$ cm, 
same as the experiment \cite{Marakov}. 
\textcolor{black}{The slip condition for superfluid and the no-slip condition 
for normal fluid are used on the walls} 
and uniform temperature \textcolor{black}{at 2 K} 
is adopted in the overall computational domain. 
Uniform distributions are applied to the superfluid and normal fluid flows 
for the inlet condition. 
\textcolor{black}{It should be noted that the hot end at $x = 0$ is not the wall 
but a location slightly away from the heated wall.} 
The convective outflow condition \textcolor{black}{\cite{outletcond}} 
is used for the outlet condition 
of the two fluids. 

\begin{figure}[t]
\centering
\includegraphics[width=0.8\linewidth]{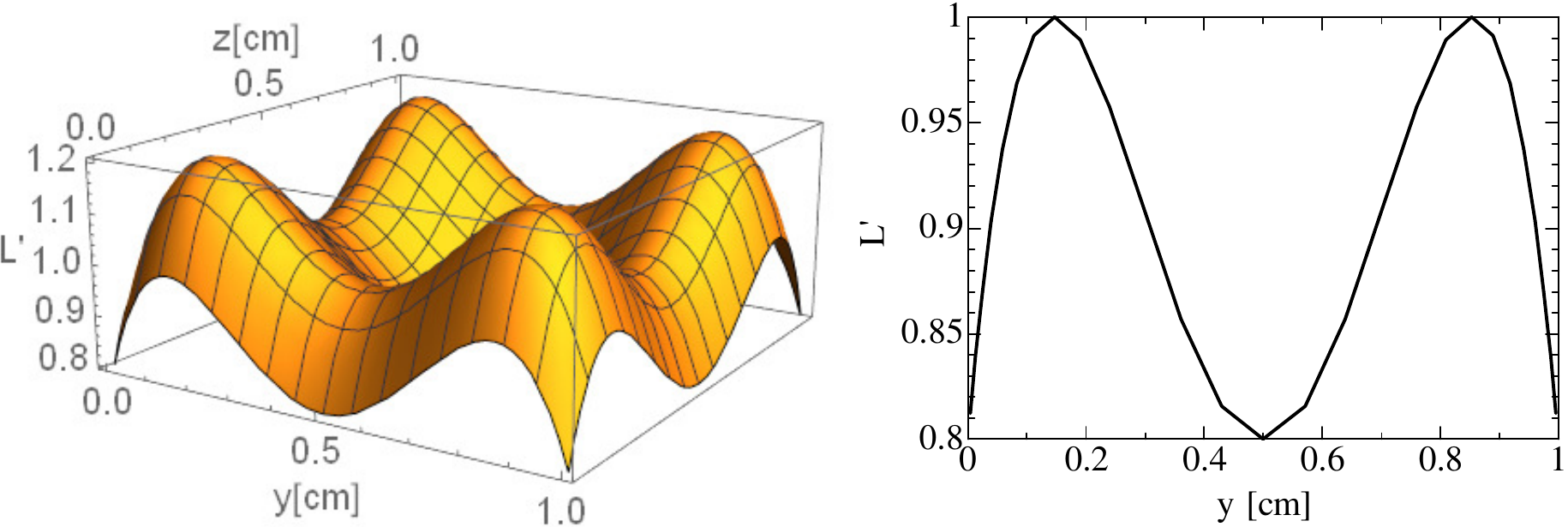}
\caption{Distribution of the vortex line density reflected to the mutual friction coefficient $a_f$ to mimic the numerical result obtained from the duct simulation by the VFM \cite{Yui-duct}. \textcolor{black}{$L'=-12.8y(y-0.5)^2(y-1)-12.8z(z-0.5)^2(z-1)+0.8$ is a non-uniform distribution normalized by the constant $a_f$ and the mean value of $L'$ is 1.01. The right figure shows the $L'$ profile in $z=0$ for the left figure.}}
\label{fig:af-dist}
\end{figure}

We carried out two conditions of mean normal fluid velocities of 
$\AV{v_n} = 0.35$ mm/s ($Re = 210$) for the parabolic flows 
and $\AV{v_n} = 3.7$ mm/s ($Re = 2227$) 
for the tail-flattened flow, where 
the Reynolds number of the normal fluid flow $Re$ is 
based on the kinematic viscosity $\nu_n=\mu_n/\rho_n$, 
the duct width $D$ and the mean normal fluid velocity. 
The grid points are $(N_x, N_y, N_z) = (481,15,15)$ 
for the parabolic flows with $a_f = 0.5, 5$ 
and $(N_x, N_y, N_z) = (31,15,15)$ 
for the tail-flattened flow with $a_f = 57$. 
The mutual friction coefficient of $a_f = 0.5$ corresponds 
to the condition of $\AV{v_n} = 0.7$ mm/s \cite{PRL-Yui}, 
and that of $a_f = 57$ is equivalent to the condition of a tail-flattened flow 
in the experiment \cite{Marakov}. 
We yielded not only a uniform $a_f$ distribution 
but also a non-uniform $a_f$ distribution due to the vortex line density 
mimicked as Fig. \ref{fig:af-dist}. 
\textcolor{black}{$L'=-12.8y(y-0.5)^2(y-1)-12.8z(z-0.5)^2(z-1)+0.8$ is 
a non-uniform distribution normalized by the constant $a_f$ 
and the mean value of $L'$ is 1.01.}
Such a non-uniform distribution has been observed 
in the VFM numerical simulation 
\cite{Baggaley-Ldist,Yui-duct} 
and the experiments \cite{Mantia-Ldist} 
of the thermal counterflows.

\vspace{-2mm}

\section{Results and Discussion}
\label{sec:3}

\vspace{-1mm}

\begin{figure}[t]
\centering
\includegraphics[width=0.95\linewidth]{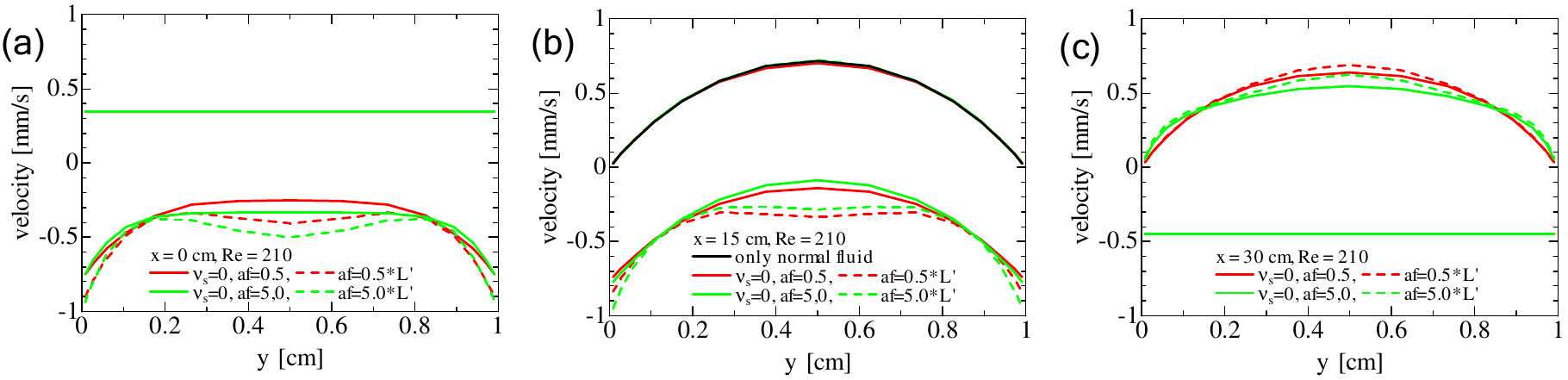}
\caption{Velocity profiles of normal fluid (positive velocity) and superfluid (negative velocity) flows at (a) $x = 0$ cm (hot end), (b) $x = 15$ cm (center of the duct) and (c) $x = 30$ cm (cold end) in the $y$ direction. Red color and green color show the mutual friction coefficient of $a_f = 0.5$ and 5. Solid line and dashed line denote the uniform and non-uniform \textcolor{black}{(multiplied by L' in Fig. 2)} distributions of the mutual friction coefficient.}
\label{fig:unus-y}
\end{figure}

Velocity profiles of the normal fluid and superfluid flows 
in the $y$ direction are shown 
in Fig. \ref{fig:unus-y}. 
A bunch of positive velocities 
and a bunch of negative velocities 
correspond to the normal fluid velocities 
and the superfluid velocities, respectively. 
We considered the effect of the non-uniform vortex line density 
\textcolor{black}{through} the non-uniform distribution 
of the mutual friction coefficient. 
The effect appears in the centerline velocity of the superfluid and normal fluid flows. 
At the hot end ($x = 0$ cm) of Fig. \ref{fig:unus-y}, 
the normal fluid velocity profiles are uniform by the inlet condition 
whereas the outlet superfluid velocity profiles become parabolic 
for uniform mutual friction distribution 
and bimodal for the non-uniform mutual friction distribution. 
These profiles are different from the BLR study \cite{PRF-Roche}. 
\textcolor{black}{Our study examines the flow regime above $T1$ transition 
in which the superfluid component transits to turbulence 
but the normal fluid is still laminar, 
whereas the BLR study mostly focuses on the flow regime below $T1$ transition 
in which both fluid components are in the laminar flow regime.} 
Therefore, we should consider $\omega_s$ directly 
or $L$ obtained from integration of $\omega$ in the mutual friction 
as mentioned in \cite{omega-const}. 

\begin{figure}[t]
\centering
\includegraphics[width=0.8\linewidth]{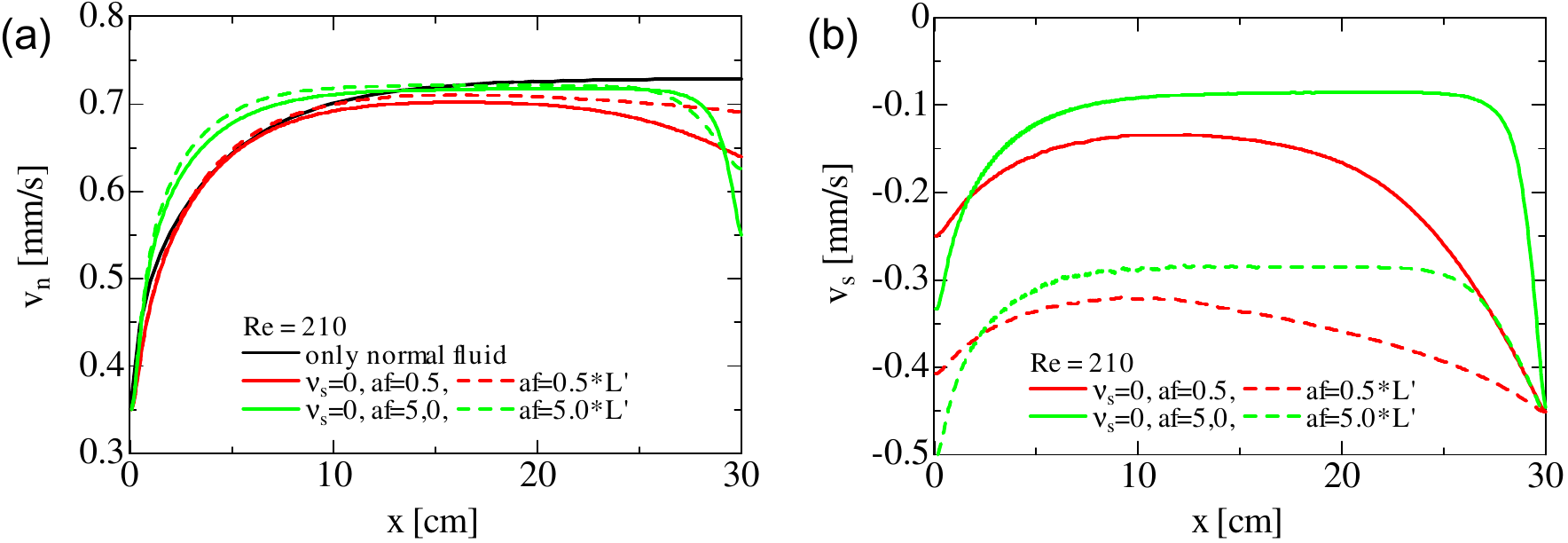}
\caption{Centerline velocity profiles of (a) normal fluid and (b) superfluid flows in the duct of $L = 30$ cm. 
\textcolor{black}{See the figure caption in Fig. \ref{fig:unus-y} for the symbols.}
}
\label{fig:unus-x}
\end{figure}

\begin{figure}[t]
\centering
\includegraphics[width=0.8\linewidth]{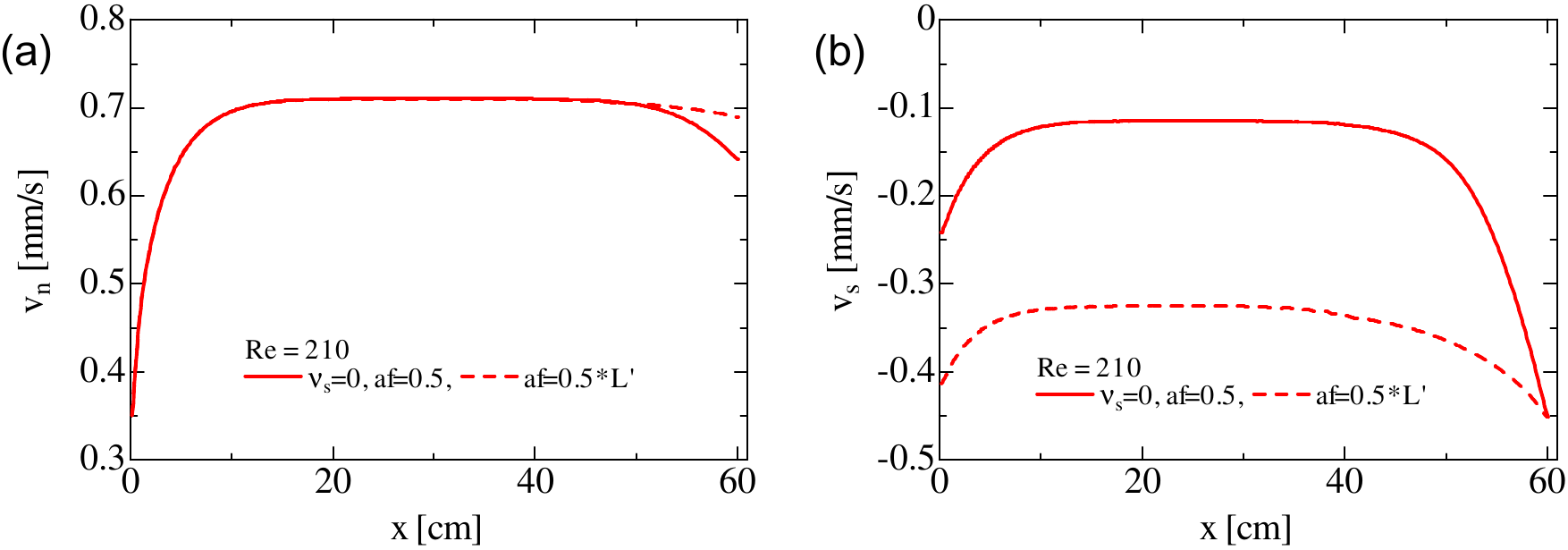}
\caption{Centerline velocity profiles of (a) normal fluid and (b) superfluid flows in the duct of $L = 60$ cm. \textcolor{black}{See the figure caption in Fig. \ref{fig:unus-y} for the symbols.}
}
\label{fig:unus-x60}
\end{figure}

Let us look at the centerline velocity profiles of the superfluid and normal fluid flows 
as shown in Fig. \ref{fig:unus-x}. 
\textcolor{black}{In the present study, the entrance length 
is measured as the length from the entrance 
where the centerline velocity in the streamwise direction 
corresponds to 99\% of maximum centerline velocity. }
The strong mutual friction shortens 
the entrance lengths of the normal fluid and superfluid flows 
at the same Reynolds number, 
e.g., $0.65X_n \rightarrow 0.59X_n$ for the normal fluid 
and $0.99X_n \rightarrow 0.31X_n$ for the superfluid 
($X_n$ denotes the entrance length of the single normal fluid), 
although in fact the mutual friction is uniquely determined 
by the relative velocity $\BM{v}_{ns}$. 
The influence of non-uniform mutual friction is slight except for the cold end 
in the normal fluid flow 
while the influence is considerable in the overall region 
of the superfluid flow. 
This is due to zero viscosity of the superfluid flow. 
It is found that 
the entrance length of the superfluid flow is longer 
than that of the normal fluid flow for weak mutual friction. 
As shown in Fig. \ref{fig:unus-x} (b), 
the centerline velocity profiles of the superfluid flow for $a_f=0.5$ 
from the cold end overlap with the profiles from the hot end, 
and thus there is a possibility that the entrance lengths for $a_f=0.5$ 
are estimated to be short. 
The examined results in a long duct of $L=60$ cm are 
displayed in Fig. \ref{fig:unus-x60}. 
The overlap of the entrance lengths is resolved in the long duct. 
It is confirmed that 
the entrance length of the superfluid flow in the short duct 
is underestimated due to the overlap, 
i.e., $0.65X_n \rightarrow 0.74X_n$ for the normal fluid 
and $0.99X_n \rightarrow 1.19X_n$ for the superfluid. 
It is also confirmed that in the superfluid flow, 
the entrance length for the non-uniform mutual friction 
becomes longer than that for the uniform mutual friction, 
i.e., $1.49X_n \leftarrow 1.19X_n$. 
\textcolor{black}{In the present study, the mutual friction parameter is assumed 
to remain constant in the streamwise direction. 
This assumption means that the vortex line density is fully developed. 
Therefore, it should be noted that 
the assumption may yield the underestimated entrance lengths 
due to preventing a feed-back mechanism to develop the vortex line density.}

\begin{figure}[t]
\centering
\includegraphics[width=0.8\linewidth]{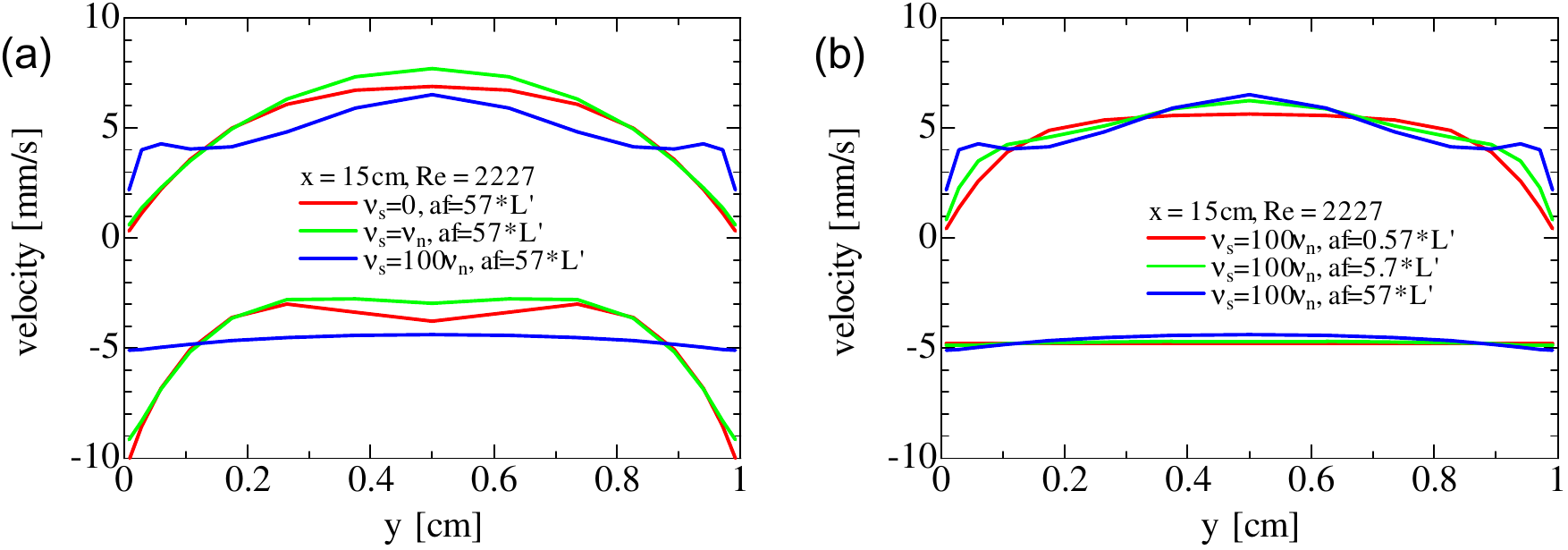}
\caption{Velocity profiles of normal fluid (positive velocity) and superfluid (negative velocity) flows with non-uniform mutual friction at $x = 15$ cm (center of the duct) in the $y$ direction; (a) the influence of the turbulent eddy viscosity of the superfluid component, (b) the influence of the mutual friction strength.}
\label{fig:unus-ytail}
\end{figure}

\textcolor{black}{A tail-flattened profile is interpreted 
as an entry effect from superfluid 
entering the duct from the helium bath in the BLR study \cite{PRF-Roche}. 
However, by using the present simulations, 
aside from the entrance effect,} 
we examined the condition to realize a tail-flattened flow 
as shown in Fig. \ref{fig:unus-ytail}. 
The non-uniform mutual friction was given to all the results 
because the uniform mutual friction showed a center-flattened flow 
and never provided the tail-flattened flow. 
\textcolor{black}{We introduce the turbulent eddy viscosity of superfluid $\nu_s$ 
that is the coefficient of the turbulent viscous term $\nu_s \nabla^2\BM{v}_s$ 
originated from the Reynolds averaged stress tensor 
of the non-linear term $\BM{v}_s\cdot \nabla \BM{v}_s$.} 
As increasing the turbulent eddy viscosity, 
the superfluid velocity profile becomes flat. 
The weak mutual friction near the walls and at the center 
yields the tail-flattened profile, 
but the large turbulent viscosity and the strong mutual friction 
are needed.

\vspace{-2mm}

\section{Conclusion}
\label{sec:4}

\vspace{-2mm}

The entrance lengths of normal fluid and superfluid flows 
in a thermal counterflow of superfluid $^4$He \textcolor{black}{at 2 K} are studied 
by using the coarse-grained HVBK two-fluid model 
for three-dimensional numerical simulations in a square duct. 
\textcolor{black}{A uniform mutual friction parameter 
was assumed in the streamwise direction.} 
It is found that the entrance length of the normal fluid flow 
from a hot end shortens when compared to that of a single normal fluid flow. 
This is due to the parabolically developed superfluid flow near the hot end 
by the mutual friction. 
The entrance length of the superfluid is longer 
than that of the normal fluid flow for weak mutual friction. 
As the mutual friction increases, the entrance lengths of the two fluids decrease 
at the same Reynolds number. 
A tail-flattened profile is realized 
by large turbulent viscosity and strong mutual friction. 

This work was supported by JSPS KAKENHI Grant Number JP18K03935. 


\vspace{-4mm}



\end{document}